\documentclass[twocolumn]{aastex631}

\usepackage{amsmath}
\usepackage{multirow}
\usepackage{hyperref}

\shorttitle{FRB population from source-count growth and burst clustering}
\shortauthors{Yamasaki et al.}

\begin{document}

\title{Fast Radio Burst Population Constrained by Source-Count Growth and Burst Clustering}

\correspondingauthor{Shotaro Yamasaki}
\email{shotaro.s.yamasaki@gmail.com}

\author[0000-0002-1688-8708]{S. Yamasaki}
\affiliation{Department of Physics, National Chung Hsing University, 145 Xingda Rd., South Dist., Taichung 40227, Taiwan}

\author[0000-0001-7228-1428]{T. Hashimoto}
\affiliation{Department of Physics, National Chung Hsing University, 145 Xingda Rd., South Dist., Taichung 40227, Taiwan}

\author[0000-0001-6010-714X]{T. Wada}
\affiliation{Department of Physics, National Chung Hsing University, 145 Xingda Rd., South Dist., Taichung 40227, Taiwan}
\affiliation{Frontier Research Institute for Interdisciplinary Sciences, Tohoku University, Sendai, Japan}
\affiliation{Astronomical Institute, Graduate School of Science, Tohoku University, Sendai, Japan}

\author[0000-0002-6821-8669]{T. Goto}
\affiliation{Institute of Astronomy, National Tsing Hua University, 101, Section 2. Kuang-Fu Road, Hsinchu, 30013, Taiwan}

\begin{abstract}
Whether all fast radio burst (FRB) sources repeat, or whether some
produce only a single burst, remains an open question. Using 3,950
bursts from the CHIME/FRB Catalog 2, we measure how the numbers of
apparently single and repeating sources change with observing time,
correcting for the declination-dependent, beam-weighted exposure and
possible variations in the detection rate. Both source counts increase
approximately linearly over 300--44,000~hr, as
$N_{\rm s}\propto\Delta t^{0.98\pm0.02}$ for single sources and
$N_{\rm r}\propto\Delta t^{1.06\pm0.14}$ for repeaters. Consequently,
the observed repeater fraction remains nearly constant at
$2.6^{+0.4}_{-0.6}\%$. We interpret these results using population
models that account for a distribution of detected burst rates and
temporal clustering. The source counts are reproduced by a population
spanning a range of rates, including sources expected to produce
approximately one detectable burst over the observing time. Its
detected burst rates follow a distribution close to $r^{-2}$ over
0.01--1.5~hr$^{-1}$. The burst timing and burst multiplicities provide
additional constraints. Among 77 repeaters, only 6 have their first two
detected bursts within a single CHIME transit, the median interval
between them is 101 days, and 31 of the 52 repeaters with at least
three bursts have five or more. These observations support this
population and favor moderately clustered emission (Weibull shape
$k\simeq0.5$) and active phases lasting months to about a year. Our
results are consistent with both a single population of repeaters whose
rate distribution steepens below the range probed by CHIME and a mixed
population that includes genuinely one-off sources. Thus, the current
data do not require a one-off population, but cannot exclude one.
Longer observing baselines are needed to distinguish these
possibilities.
\end{abstract}

\keywords{Radio transient sources (2008) --- Fast radio bursts (2008) --- Transient sources (1851)}

\section{Introduction}
\label{sec:intro}

Fast radio bursts (FRBs) are millisecond radio transients of
extragalactic origin \citep{lorimer07,petroff22}. Their sources are
classified as apparently one-off or repeating. This classification is
incomplete, because within a finite observing period one cannot tell
whether an apparently one-off source will burst again
\citep{caleb19,james20a}. In the CHIME surveys only a few per cent of
the sources are classified as repeaters
\citep{chime-catalog1,chime-repeaters23,chime-catalog2,cook26}. The
true fraction matters for the nature of FRBs. Genuinely one-off sources
could come from cataclysmic progenitors, whereas repeaters require
persistent engines. Population studies suggest that many FRBs may
ultimately repeat \citep[e.g.,][]{ravi19,james23,beniamini25,cook26,hoffmann26},
while repeaters and apparently one-off sources differ in burst
morphology \citep[e.g.,][]{pleunis21,chime-catalog2}.

The growth of the numbers of single and repeating sources with
observing time offers a direct probe \citep{ai21}. With the first
CHIME/FRB catalog, \citet[hereafter Y24]{yamasaki24} and
\citet{mcgregor24} found that most FRBs may repeat. If apparently
one-off sources are repeaters that have not yet repeated, this growth
reflects the distribution of burst rates and the clustering of bursts.

Here we measure both source counts (SCs) over the 5-yr time baseline of the
second CHIME/FRB catalog. The SCs constrain the rate distribution but
hardly the clustering. We therefore add two properties of the
repeaters, the timing of their first two bursts and the number of their
detected bursts. With both, we ask whether the data require genuinely
one-off FRBs.

This {\it Letter} is organized as follows. \S~\ref{sec:sc_measurement}
describes the sample and the source-count construction.
\S~\ref{sec:sc_models} models the source counts with distributions of
burst rates. \S~\ref{sec:timing} uses the repeat bursts of the repeaters
to constrain the clustering. \S~\ref{sec:discussion} discusses the
physical implications and future tests, and \S~\ref{sec:conclusions}
summarizes.

\section{Source counts}
\label{sec:sc_measurement}

We use the second CHIME/FRB catalog (Cat2; \citealt{chime-catalog2}),
from 2018 September 4 to 2023 September 15, a survey of 1,837 days. Of
the 5,023 entries in this period, we remove those flagged as unsuitable
for population analyses, which leaves 4,628. Requiring ${\rm S/N}>10$
and a measured declination leaves 4,409. Cat2 lists the sub-bursts of a
burst as separate entries that share one event identifier. We keep one
entry per event, so that a sub-burst is neither counted as a new single
source nor taken as a repeat of its own burst. The final sample has
3,950 bursts, 3,076 from single sources and 874 from 81 repeaters
identified by CHIME. We use the CHIME associations only to group bursts
into sources, and classify each source by the number of its bursts that
pass our cuts.

The flagged events fall on 152 of the 1,837 days, and we exclude these
days from the SC construction. They remove whole daily cycles of
exposure, and because the SC is cumulative, the missing growth would
lower all later counts. For each excluded interval we therefore add the
growth expected from the mean rate over the preceding 100~hr, separately
for singles and repeaters, and carry it to all later epochs. This is a
local linear interpolation within each gap and adds nothing outside the
excluded days. Without this correction the growth indices
below change by less than 0.03 and the repeater-to-single ratio by less
than 0.001.

\begin{figure}
\centering
\includegraphics[width=\columnwidth]{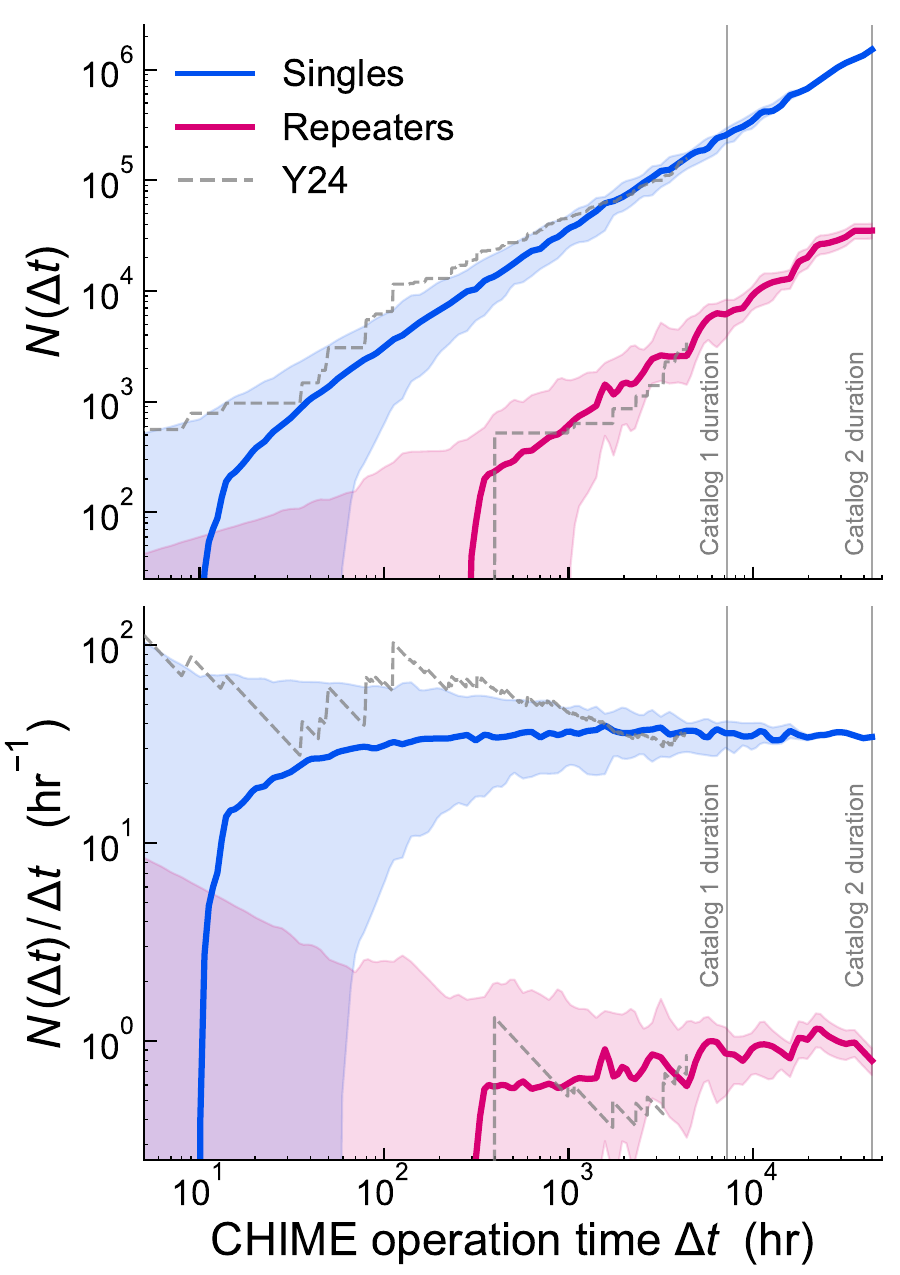}
\caption{Top: Exposure-corrected SCs of single (blue) and repeating
(pink) sources versus operation time $\Delta t$. Solid lines are the
medians over independent intervals of duration $\Delta t$, and shaded
regions their uncertainties. The grey dashed curve is the SC of a single
interval within the Cat1 period, as in Y24. Grey vertical lines mark the
durations of Catalogs 1 and 2. Bottom: Average rate $N/\Delta t$.}
\label{fig:sc_measured}
\end{figure}

CHIME observes high-declination sources much longer than those near its
southern limit, which would bias the apparent repeater fraction. We
weight each source $i$ by its exposure (Y24),
\begin{equation}
f_i=\frac{T_{\rm eff}(\delta_i)}{T_{\rm survey}},
\label{eq:exposure_weight}
\end{equation}
where $T_{\rm eff}(\delta)=\sum_jT_jB_j^{3/2}$ sums the time along the
transit weighted by the beam sensitivity $B$ relative to the beam
center, averaged over 400--800~MHz \citep{james23}. The power $3/2$ is
the fraction of bursts above threshold for a Euclidean fluence
distribution, which Cat2 follows above $\simeq10$~Jy~ms. We fix
$T_{\rm survey}=1,837\times0.55$~d by reproducing the catalog exposure
(Appendix~\ref{app:exposure}). The exposure-corrected SC after an
elapsed observing time $\Delta t$, which we call the time baseline, is
\begin{equation}
N(\Delta t)=\sum_{t_i<\Delta t}\frac{1}{f_i},
\label{eq:source_count}
\end{equation}
where $t_i$ is the elapsed time from the start of the observation to the
first detected burst from source $i$. With $f_i=1$,
Eq.~\eqref{eq:source_count} gives the raw count $n(\Delta t)$. A source
is a single at its first burst and becomes a repeater at its second.

To include variations of the detection rate, we divide the survey into
all non-overlapping intervals of length $\Delta t$, build the SC in each,
and take the median. Its uncertainty is the larger of the scatter
between intervals and the Poisson and exposure errors. We evaluate the
SC on a logarithmic grid of $\Delta t$ from 1~hr to 44,088~hr, with 20
values per decade. The
uncertainty at each $\Delta t$ sets the weights of the fits. All
time baselines use the same bursts and are therefore correlated, so we
estimate the uncertainties of the fitted quantities with a bootstrap
over sources. We resample the 3,157 sources 1,000 times, keeping all
bursts of a repeater together, and repeat the whole construction and
fit.

Figure~\ref{fig:sc_measured} shows the result. Both counts grow
steadily from $\Delta t\sim300$~hr to the end of the survey. On 44
logarithmically spaced time baselines over
$300~{\rm hr}\leq\Delta t\leq4.4\times10^4$~hr we use
three diagnostics. The growth indices are
\begin{equation}
\beta_s\equiv\frac{d\ln N_s}{d\ln\Delta t}=0.98\pm0.02,\quad
\beta_r\equiv\frac{d\ln N_r}{d\ln\Delta t}=1.06\pm0.14,
\label{eq:growth_slopes}
\end{equation}
the count--count slope, which does not involve the time axis, is
\begin{equation}
\Gamma_{\rm rs}\equiv\frac{d\ln N_r}{d\ln N_s}=1.08^{+0.14}_{-0.16},
\label{eq:count_slope}
\end{equation}
and the repeater-to-single ratio,
\begin{equation}
Q\equiv\frac{N_r}{N_s}=0.026^{+0.004}_{-0.006},
\label{eq:ratio}
\end{equation}
is consistent with being constant. Both counts grow linearly with no
break. A single interval of the Cat1 length, as used in Y24, scatters
strongly at $\Delta t\lesssim10^3$~hr and is not representative. The
last year of the survey contains fewer new repeaters than the earlier
years. Ending the sample one year earlier, on 2022 September 15, raises
$\beta_r$ and $\Gamma_{\rm rs}$ by about 0.1, within $1\sigma$.
Table~\ref{tab:observables} lists these diagnostics, used in
\S~\ref{sec:sc_models}, together with the properties of the repeat
bursts used in \S~\ref{sec:timing}.

\begin{deluxetable}{lll}
\tablecaption{Quantities used to constrain the models.
\label{tab:observables}}
\tablewidth{0pt}
\tabletypesize{\footnotesize}
\tablehead{
\colhead{Quantity} & \colhead{Definition} & \colhead{Value}}
\startdata
$\Gamma_{\rm rs}$ & $d\ln N_r/d\ln N_s$ & $1.08^{+0.14}_{-0.16}$\\
$\beta_s$ & $d\ln N_s/d\ln\Delta t$ & $0.98\pm0.02$\\
$\beta_r$ & $d\ln N_r/d\ln\Delta t$ & $1.06\pm0.14$\\
$Q$ & $N_r/N_s$ & $0.026^{+0.004}_{-0.006}$\\\hline
$F_{\rm same}$ & first repeats within one transit & $6/77=0.078^{+0.036}_{-0.025}$\\
$\tau_{1/2}$ & median first-repeat interval & 101~d\\
$F_{5|3}$ & $\geq5$ bursts among $\geq3$ & $31/52=0.60^{+0.06}_{-0.07}$\\
\enddata
\tablecomments{The first four are the SC diagnostics, with uncertainties
from a bootstrap over sources. The last three describe the repeat
bursts (\S~\ref{sec:timing}), with 68\% binomial intervals.}
\end{deluxetable}

\section{Rate distribution from the source counts}
\label{sec:sc_models}

\subsection{Formulation}
\label{sec:formulation}

A source with detected burst rate $r$ at declination $\delta$ expects
\begin{equation}
\mu\equiv rf\Delta t
\label{eq:mu}
\end{equation}
detections over $\Delta t$, where $r$ is the rate of bursts detectable
on the beam axis. The exposure is a union of short transits,
$W(\Delta t)$, once per sidereal day, with a lower transit 12~hr later
for circumpolar sources ($\delta\geq70^\circ$), and its total duration
is $|W(\Delta t)|=f\Delta t$. We define the probabilities of zero, one,
and two or more detections in these windows,
\begin{equation}
P_i^{\rm win}(\Delta t\,|\,r,\delta)\equiv\Pr\left[N(W(\Delta t))= i\right],
\quad i\in\{0,1,\geq2\}.
\label{eq:p_win}
\end{equation}
Within a transit, a burst is detected with probability $p=B^{3/2}$,
which integrates to the same exposure as in the counts. For $N_{\rm tot}$
sources with rate distribution $\phi(r)$, the exposure-corrected counts
are
\begin{equation}
N_{s,r}(\Delta t)=N_{\rm tot}\int dr\,\phi(r)
\left\langle f^{-1}P_{1,\geq2}^{\rm win}\right\rangle_\delta,
\label{eq:expected_counts}
\end{equation}
where $\langle\cdot\rangle_\delta$ denotes the average over source
declinations for an isotropic sky. For Poisson bursts,
$P_1=\mu e^{-\mu}$ and $P_{\geq2}=1-e^{-\mu}(1+\mu)$ for any window
arrangement. For clustered bursts we evaluate the probabilities by Monte
Carlo with the actual transits (Appendix~\ref{app:monte_carlo}).

A source with $\mu\ll1$ is rarely detected and is then almost always a
single. A source with $\mu\gg1$ is a repeater. Only sources with
$\mu\sim1$ move from singles to repeaters as $\Delta t$ grows. With
$\langle f\rangle_\delta\simeq0.0022$, the survey probes them over
\begin{equation}
1.0\times10^{-2}~{\rm hr}^{-1}\lesssim r\lesssim1.5~{\rm hr}^{-1}.
\label{eq:probed_range_value}
\end{equation}
This range is approximate because $f$ varies with declination. We first
consider populations that lie entirely below or above it, and then
populations that span it.

\subsection{Constraints on the rate distribution}
\label{sec:regimes}

\paragraph{All sources with $\mu\gg1$.} Every source is a
repeater, so $N_s\to0$ and $N_r$ stops growing, for any clustering.
This is excluded.

\paragraph{All sources with $\mu\ll1$.} In this regime
$P_1\simeq\mu$ for any $\phi(r)$ and any clustering, so
$N_s\propto\Delta t$, as observed. For Poisson bursts
$P_{\geq2}\simeq\mu^2/2$, so $N_r\propto\Delta t^2$ and
$\Gamma_{\rm rs}\simeq2$, far above the $2\sigma$ limit of 1.37. For
clustered bursts we use stationary Weibull clustering of shape $k$, a
renewal process with Weibull waiting times that reduces to Poisson at
$k=1$. At low rate a repeat needs a pair of bursts, and clustering makes
pairs within one transit common, $P_{\geq2}\propto\mu^{1+k}$ in one
window. Through the daily transits, $N_r$ is set by a pair factor
$\Pi_k(\Delta t)$ that weights pairs of observed instants by their
separation as $\tau^{k-1}$ (Appendix~\ref{app:low_rate_pairs}). Pairs
within a transit grow as $\Delta t$ and pairs in different transits as
$\Delta t^{1+k}$. With beam-weighted transits, $\Gamma_{\rm rs}$ stays
within $2\sigma$ for $k\lesssim0.54$ (Figure~\ref{fig:pair_factor_slope}).
These clustered low-$\mu$ populations remain and are tested in
\S~\ref{sec:timing}.

\paragraph{Populations spanning $\mu\sim1$.} A population spanning the
transition, with $\mu\sim1$ for the sources that produce the repeaters,
needs a broad rate distribution. The simplest one has no preferred
scale. Since the probabilities depend on $r$ and $\Delta t$ mainly
through $\mu=rf\Delta t$ (only through $\mu$ for Poisson bursts), a
distribution whose shape is invariant under $r\to cr$ must be a power
law (PL),
\begin{equation}
\frac{dN}{dr}=Cr^{-\alpha},\qquad r_{\min}\leq r\leq r_{\max}.
\label{eq:power_law}
\end{equation}
For Poisson bursts with $r_{\min}\to0$ and $r_{\max}\to\infty$, changing
the integration variable in Eq.~\eqref{eq:expected_counts} from $r$ to
$\mu$ gives $N_s,N_r\propto\Delta t^{\alpha-1}$, so that
\begin{equation}
\beta_s=\beta_r=\alpha-1,\qquad
Q=\frac{2-\alpha}{\alpha-1}.
\label{eq:power_law_scaling}
\end{equation}
The repeater integral converges for $1<\alpha<3$, so $\beta_r=\alpha-1$
holds there. The singles integral also needs $\alpha<2$. For
$\alpha\geq2$ it is dominated by the lowest rates, so $N_s\propto\Delta t$
and $Q$ depends on $r_{\min}$. For $1<\alpha<2$ the count--count slope
is exactly unity and $Q$ is constant, as observed. The measured
$\beta_r$, $\beta_s$, and $Q$ give $\alpha=2.06\pm0.14$, $1.98\pm0.02$,
and $1.975^{+0.005}_{-0.004}$, so a scale-free population with
$\alpha\simeq2$ reproduces all three diagnostics. Below we use
$\alpha=2.06$, the value from $\beta_r$, as the fiducial index.

This limit, however, has $N_{\rm tot}\to\infty$, with almost all sources
never bursting, so a real population needs a finite $r_{\min}$. Sources
below the probed rates add singles but almost no repeaters, so the SCs
fix only how many such sources there are, through $Q$. Removing the
lowest-rate sources raises $Q$, so for $\alpha<1.975$ even the
scale-free $Q$ exceeds the observed value, and no cutoff helps. For
$\alpha>2$ the singles grow as $r_{\min}$ decreases, and $Q$ is matched
at a finite cutoff. Matching it at a reference time baseline of 3981~hr, for
Poisson bursts this needs recurrence times
$1/r_{\min}\sim7\times10^{13}$, $4\times10^3$, and 30~yr for
$\alpha=2.0$, 2.1, and 2.2. The diagnostics stay within $1\sigma$ for
$\alpha\leq2.15$. Clustering adds repeaters, so it needs an even lower
cutoff. If the sources are magnetars, whose active ages are
$\sim10^4$~yr \citep{kaspi17,beniamini19}, a cutoff within them needs
$\alpha\gtrsim2.14$ with the clustering of \S~\ref{sec:timing}. For the
fiducial $\alpha=2.06$, the single PL thus needs a cutoff far below the
probed rates and implies an enormous number of sources that almost
never burst.

\subsection{Relaxing the low-rate extrapolation}
\label{sec:relaxing}

The data do not require the PL to continue far below the probed rates.
Two simple forms supply the singles required by $Q$ with a cutoff close
to the probed rates, for any $\alpha$ near 2. One steepens the low-rate
tail of the repeating population, and the other adds a genuinely
one-off population.

\begin{figure*}
\centering
\includegraphics[width=\textwidth]{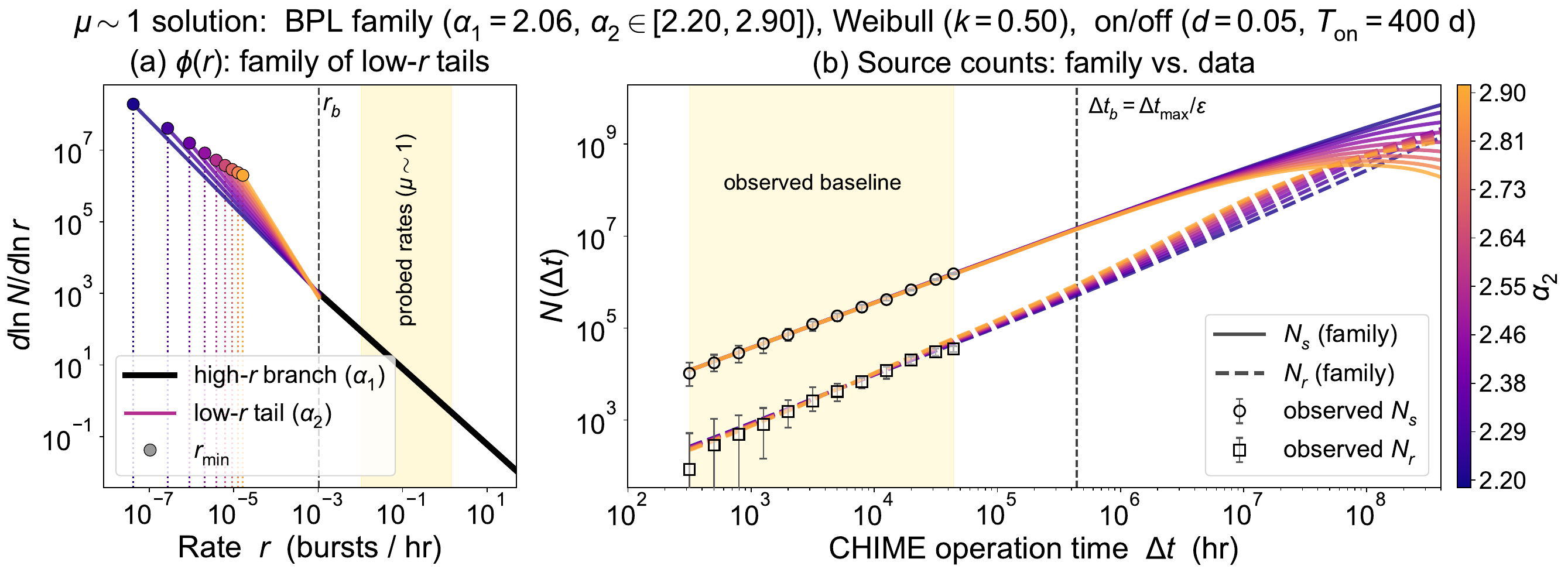}
\caption{BPL family ($\alpha_1=2.06$, $\alpha_2=2.2$--2.9, color) with
on/off activity (Weibull clustering of shape $k=0.5$ within active
phases of mean duration $T_{\rm on}=400$~d and duty cycle $d=0.05$),
observed through beam-weighted transits. (a) Rate distribution. The
high-rate branch (black) is shared by all members, and each low-rate
tail ends at its $r_{\min}$ (dots). The shaded band is the probed range
(Eq.~\ref{eq:probed_range_value}), and the vertical line marks the break
rate $r_b$. (b) $N_s$ (solid) and $N_r$ (dashed) compared with the data,
and extended to future time baselines. The vertical line marks
$\Delta t_b=\Delta t_{\max}/\epsilon$, where sources near the break start
to repeat.}
\label{fig:bpl_family}
\end{figure*}

\paragraph{A broken PL (BPL).} The PL of Eq.~\eqref{eq:power_law} is
kept at and above a break $r_b$, now with index $\alpha_1$, and below
the break the tail steepens,
\begin{equation}
\frac{dN}{dr}\propto
\begin{cases}
r^{-\alpha_2}, & r_{\min}\leq r<r_b,\\[2pt]
r^{-\alpha_1}, & r_b\leq r\leq r_{\max},
\end{cases}
\qquad \alpha_2>\alpha_1,
\label{eq:bpl}
\end{equation}
where the two parts join at $r_b$. We call the two parts the low-rate tail and the
high-rate branch. We place the break
below the probed rates, $r_b=\epsilon/(\langle f\rangle_\delta\Delta
t_{\max})$ with $\epsilon=0.1$, where $\Delta t_{\max}=4.4\times10^4$~hr is the longest time baseline, so that sources below the break have
$\mu\leq\epsilon\ll1$. For $\alpha_2>2$ the tail adds singles without bound
as $r_{\min}$ decreases, so $Q$ is matched at a finite cutoff, while the
tail adds few repeaters. Figure~\ref{fig:bpl_family} shows the family
with the fiducial $\alpha_1=2.06$ and $\alpha_2=2.2$--2.9, computed with the clustering that the repeat bursts
require (\S~\ref{sec:onoff}). The SCs are almost insensitive to this
clustering. Each member matches $Q$ with a cutoff recurrence time of
$3\times10^3$ to 7~yr, and all reproduce the SCs within $1\sigma$
($\beta_s=0.97$--0.98, $\beta_r=1.05$--1.17, $\Gamma_{\rm rs}=1.07$--1.21).
The SCs determine neither $\alpha_2$ nor $r_b$.

\paragraph{A genuinely one-off population.} The repeaters keep a single
PL (Eq.~\ref{eq:power_law}), and a one-off component, such as compact binary mergers, adds
singles at a constant rate (\citealt{ai21}, Y24). We define its fraction
$f_{\rm nonrep}$ of the singles at the reference time baseline. With
$\alpha=2.06$ and the same clustering, members with
$f_{\rm nonrep}=0.01$--0.95 all reproduce the SCs within $1\sigma$, and a
cutoff within the magnetar ages needs $f_{\rm nonrep}\gtrsim0.45$
(Appendix~\ref{app:two_population}).

The SCs thus leave two classes of model, clustered low-$\mu$ populations
with $k\lesssim0.54$ and populations spanning $\mu\sim1$ with
$\phi(r)\propto r^{-2}$ over the probed rates. For the latter, a single
PL, a BPL, and a two-population model all remain.

\section{Constraints from the repeat bursts}
\label{sec:timing}

\subsection{Repeat bursts in the catalog}
\label{sec:first_repeat}

\begin{figure*}
\centering
\includegraphics[width=\textwidth]{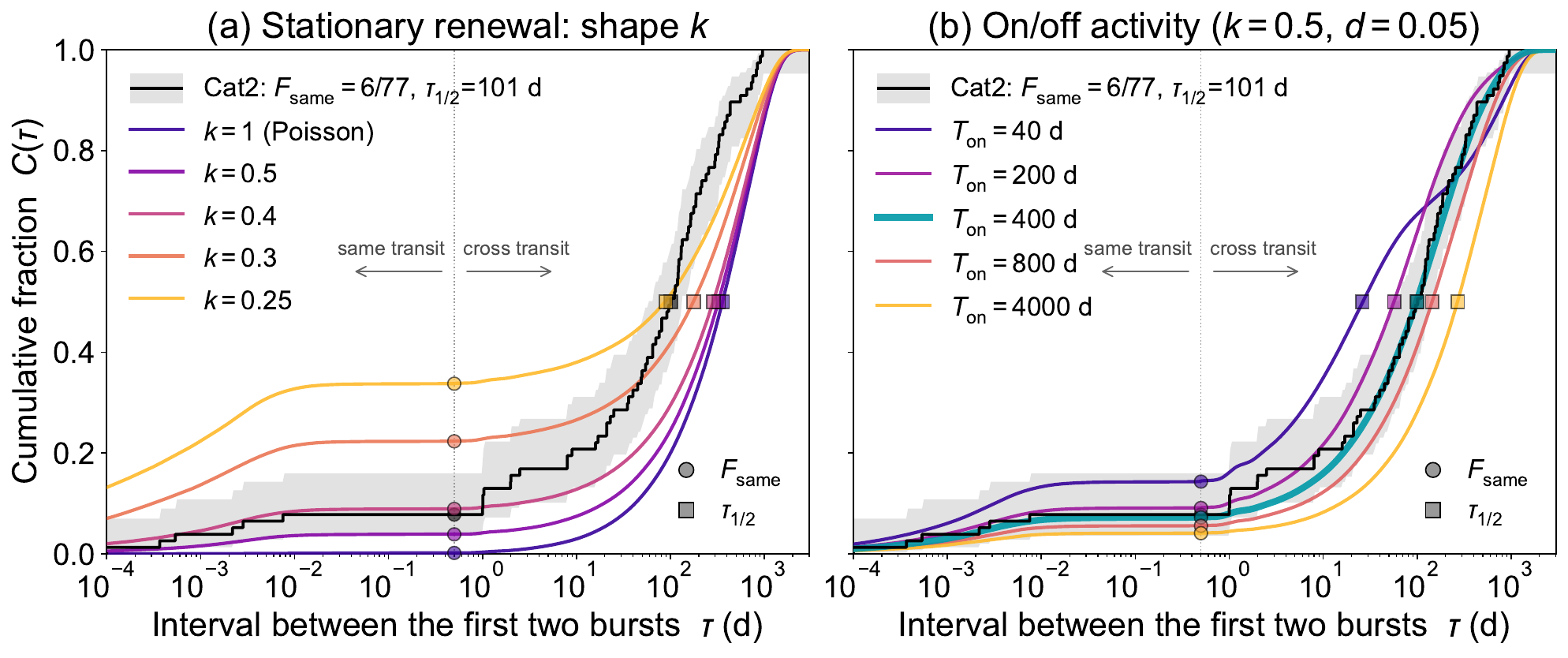}
\caption{Cumulative distribution $C(\tau)$ of the interval between the
first two detected bursts of the 77 catalog repeaters (black, with the
95\% interval in gray), compared with the BPL ($\alpha_1=2.06$,
$\alpha_2=2.2$) observed through beam-weighted transits. The dotted line
marks half a day, where $C=F_{\rm same}$. (a) Stationary Weibull
clustering with shape $k=0.25$--1 ($k=1$ is Poisson). (b) On/off
activity with $k=0.5$ and $d=0.05$, for mean active durations
$T_{\rm on}=40$--4000~d. The thick line is the representative model,
$T_{\rm on}=400$~d. Circles mark $F_{\rm same}$ and squares mark
$\tau_{1/2}$.}
\label{fig:first_repeat_intervals}
\end{figure*}

After our cuts (\S~\ref{sec:sc_measurement}), 77 of the 81 repeaters
identified by CHIME have at least two events. The other four have only
one event left and count as singles. We use the interval $\tau$ between
the first two detected bursts from these 77 repeaters, and its
cumulative distribution $C(\tau)$, the fraction of repeaters with an
interval shorter than $\tau$ (black lines in
Figure~\ref{fig:first_repeat_intervals}). For 6 repeaters the two bursts
fall in one transit, so
\begin{equation}
F_{\rm same}\equiv C(0.5\,{\rm d})=\frac{6}{77}=0.078,
\label{eq:fsame_obs}
\end{equation}
with a 95\% interval of 0.04--0.16. Half a day is the shortest time
between two transits of a source, the upper and lower transits of a
circumpolar source, so $F_{\rm same}$ counts the first repeats within one
transit. Beyond a day the distribution is broad, with a median
$\tau_{1/2}=101$~d. Figure~\ref{fig:first_repeat_intervals} marks
$F_{\rm same}$ and $\tau_{1/2}$ with circles and squares.

We also use the number of detected bursts. Of the 77 repeaters, 52 have
three or more events, and 31 of these have five or more. We use
\begin{equation}
F_{5|3}=\frac{31}{52}=0.60,
\label{eq:f53_obs}
\end{equation}
with a 95\% interval of 0.46--0.72. Repeaters with three or more bursts
are identified reliably, whereas those with only two are harder to
separate from a chance association \citep{cook26}. As we show in
\S~\ref{sec:multiplicity}, thresholds of four or ten bursts,
$F_{4|3}$ and $F_{10|3}$, give the same constraints on the models.

For each model we simulate the bursts of sources at each rate through
the transits and record their detections. $F_{\rm same}$ and $F_{5|3}$
follow by weighting each rate by $\phi(r)$ times the probability of the
relevant detections, without the factor $f^{-1}$, as in the catalog. We
compare them with the catalog through the binomial probability, and
$C(\tau)$ at $\tau>1$~d through the Kolmogorov--Smirnov (KS) test.

\subsection{Stationary clustering fails}
\label{sec:timing_test}

We first test stationary Weibull clustering. The low-$\mu$ populations
allowed by the count--count slope match $F_{\rm same}$, but their median first-repeat interval is three to four
times too long ($\tau_{1/2}=300$--430~d, KS probability $<10^{-9}$). For
populations spanning $\mu\sim1$, Figure~\ref{fig:first_repeat_intervals}a
shows the intervals predicted for $k=0.25$--1. The shape $k$ sets the
step at short intervals, that is, $F_{\rm same}$. Poisson bursts ($k=1$)
give $F_{\rm same}\simeq0.002$, far below the catalog, and the observed
value requires $k\simeq0.42$ (68\% range 0.37--0.46). The longer
intervals, however, hardly change with $k$. For $k=0.4$--1, $\tau_{1/2}=290$--360~d,
about three times the observed value (KS probability $<10^{-7}$).
Stationary clustering links bursts only over a few mean intervals, while
CHIME sees a source only a small fraction of the time. Detections on
different days are then nearly independent and fall at random within the
survey. Both classes therefore fail.

\subsection{On/off activity}
\label{sec:onoff}

A second, longer clustering timescale shortens the intervals. We
consider a toy model in which each source alternates between active and
quiescent phases. The length of each phase is drawn from an exponential
distribution with mean $T_{\rm on}$ for active phases and
$T_{\rm off}=T_{\rm on}(1-d)/d$ for quiescent ones, where $d$ is the duty
cycle. While active, the bursts follow Weibull clustering of shape $k$
at rate $r/d$, so the mean rate stays $r$. The exponential distribution
is the simplest choice that fixes only the mean durations. The actual
distribution is poorly known. Some repeaters show periodic activity
windows \citep{chime20_periodic,rajwade20,cruces21}, whereas others show
single episodes of high activity lasting months \citep{lanman22}. We
also test fixed durations below. Figure~\ref{fig:first_repeat_intervals}b
shows that $T_{\rm on}$ sets the long intervals.

A likelihood over $k=0.35$--0.65, $d=0.01$--0.3, and
$T_{\rm on}=100$--1200~d, combining $F_{\rm same}$ with each interval
longer than a day, gives the 68\% ranges $k=0.50^{+0.04}_{-0.05}$ and
$T_{\rm on}=400^{+190}_{-70}$~d for the BPL, and
$k=0.50^{+0.03}_{-0.08}$ and $T_{\rm on}=400^{+270}_{-70}$~d for the
two-population model. Both give $d\lesssim0.07$, and $d$ up to about 0.2
is allowed at 95\%. One cycle, $T_{\rm on}/d$, lasts more than about 5~yr. This is as long as the longest time baseline, $\Delta t_{\max}=4.4\times10^4$~hr, so the intervals measure how long a source stays active, not whether it becomes active again. We adopt $k=0.5$, $d=0.05$, and $T_{\rm on}=400$~d as a
representative model. For the BPL it gives $F_{\rm same}=0.072$ and
$\tau_{1/2}=99$~d, with a KS probability of 0.90. Fixed on/off durations
(strictly periodic activity) need $T_{\rm on}\simeq1000$--1400~d and fit
slightly worse, by $\Delta(-2\ln L)=3.2$ and 2.7.

On/off activity barely changes the SCs. Because one cycle is longer than $\Delta t_{\max}$, a source has at most about one active phase within any time baseline. On time baselines shorter than
$T_{\rm on}$ the population looks like a fraction $d$ of the sources
bursting at rate $r/d$, which scales both counts by $d^{2-\alpha}$. On longer time baselines, most active phases begin and end within $\Delta t$. A source has at most one such phase, so each phase is seen as a separate source. The number of these phases grows as $d\Delta t/T_{\rm on}$, while each contains on average $rT_{\rm on}/d$ bursts, independent of $\Delta t$. This scales both counts by $(d\Delta t/T_{\rm on})^{2-\alpha}$.
For
$\alpha=2$ the counts are unchanged in both cases, because such a distribution has the
same number of bursts in every logarithmic interval of rate. For $\alpha$ near 2 the factor changes only slowly with $\Delta t$, and $\Gamma_{\rm rs}$ barely changes because both counts scale by the same factor. The
families of Figure~\ref{fig:bpl_family} and
Appendix~\ref{app:two_population} give slopes within $\sim0.03$ of the
stationary models, except the steepest BPL tails, whose count--count slope is
up to 0.08 higher.

On/off activity also shortens the intervals of the low-$\mu$ models. It
packs their bursts into active phases, however, which adds pairs in
different transits and steepens $N_r$. $\Gamma_{\rm rs}$ then stays within
$2\sigma$ only for $\mu\lesssim0.003$. Such populations with
$k\simeq0.5$--0.6 reproduce $F_{\rm same}$ and the intervals, so we test
them with the number of bursts.

\subsection{The number of bursts per repeater}
\label{sec:multiplicity}

The low-$\mu$ populations fail $F_{5|3}$. In them a repeater is a rare
pair of detections, and more detections are rarer still. They give
$F_{5|3}<10^{-3}$, against 0.60, while the catalog has 14 repeaters with
ten or more bursts. Thresholds of four or ten bursts give the same
result, with $F_{4|3}\leq0.16$ and $F_{10|3}<10^{-3}$ against the
observed $37/52=0.71$ and $14/52=0.27$. Moreover, the repeaters with many bursts
reach $\mu\gtrsim1$, at least while active, and the low-$\mu$
populations are excluded.

The $\mu\sim1$ models with on/off activity give $F_{5|3}=0.46$ for the
single PL with $\alpha=2.06$ and for the two-population model, and 0.45
for the representative BPL, all within $2\sigma$. A steeper rate
distribution adds sources detected only three or four times and lowers
$F_{5|3}$. The single PL with $\alpha=2.3$ and the BPL with
$\alpha_2=2.5$ give 0.37 and 0.42, $3.2\sigma$ and $2.5\sigma$ below the
catalog. The number of bursts thus mildly favors the flatter members of
the models allowed by the SCs, $\alpha\lesssim2.1$ and
$\alpha_2\lesssim2.2$. With four or ten bursts instead of five, the BPL
with $\alpha_2=2.5$ also lies within $2\sigma$. The fraction of all
repeaters with three or more bursts is 0.68 in the catalog and about 0.45
in these models. We do not use it, because it depends on how many
repeaters with only two bursts are identified.

\begin{deluxetable*}{llllccccccl}
\tablecaption{Summary of the model constraints.
\label{tab:model_summary}}
\tablewidth{0pt}
\tabletypesize{\scriptsize}
\tablehead{
\colhead{$\mu$ regime} & \colhead{$\phi(r)$} & \colhead{Clustering} &
\colhead{$\Gamma_{\rm rs}$} & \colhead{$\beta_s$} & \colhead{$\beta_r$} &
\colhead{$Q$} & \colhead{$F_{\rm same}$} & \colhead{$\tau_{1/2}$} & \colhead{$F_{5|3}$} & \colhead{Remark}}
\startdata
\hline
\multirow{3}{*}{$\mu\ll1$} & \multirow{3}{*}{Any} & $k\gtrsim0.54$ & $\times$ & $\checkmark$ & $\times$ & -- & -- & -- & -- & Excluded by $\Gamma_{\rm rs}$ (Fig.~\ref{fig:pair_factor_slope})\\
 & & $k\lesssim0.54$ & $\checkmark$ & $\checkmark$ & $\checkmark$ & $\checkmark$ & $\checkmark$ & $\times$ & -- & Intervals too long (\S~\ref{sec:timing_test})\\
 & & On/off, $k\simeq0.5$ & Edge & $\checkmark$ & $\checkmark$ & $\checkmark$ & Edge & $\checkmark$ & $\times$ & Too few bursts per repeater (\S~\ref{sec:multiplicity})\\
\hline
$\mu\gg1$ & Any & Any & $\times$ & $\times$ & $\times$ & $\times$ & -- & -- & -- & Singles vanish (\S~\ref{sec:sc_models})\\
\hline
\multirow{10}{*}{$\mu\sim1$} & PL, $r_{\min}\to0$ & Poisson & $\checkmark$ & $\checkmark$ & $\checkmark$ & $\checkmark$ & $\times$ & $\times$ & -- & $\alpha\simeq1.97$--1.98, but $N_{\rm tot}\to\infty$\\
 \cline{2-11}
 & \multirow{3}{*}{PL, finite $r_{\min}$} & Poisson & $\checkmark$ & $\checkmark$ & $\checkmark$ & $\checkmark$ & $\times$ & $\times$ & -- & $\alpha\simeq2$--2.15, timing fails\\
 & & $k\simeq0.42$ & $\checkmark$ & $\checkmark$ & $\checkmark$ & $\checkmark$ & $\checkmark$ & $\times$ & -- & Intervals too long\\
 & & On/off, $k\simeq0.5$ & $\checkmark$ & $\checkmark$ & $\checkmark$ & $\checkmark$ & $\checkmark$ & $\checkmark$ & $\checkmark$ & Viable for $\alpha\simeq2$--2.1\\
 \cline{2-11}
 & \multirow{3}{*}{BPL, $\alpha_1=2.06$, $\alpha_2=2.2$--2.9} & Poisson & $\checkmark$ & $\checkmark$ & $\checkmark$ & $\checkmark$ & $\times$ & $\times$ & -- & Timing fails\\
 & & $k\simeq0.42$ & $\checkmark$ & $\checkmark$ & $\checkmark$ & $\checkmark$ & $\checkmark$ & $\times$ & -- & Intervals too long\\
 & & On/off, $k\simeq0.5$ & $\checkmark$ & $\checkmark$ & $\checkmark$ & $\checkmark$ & $\checkmark$ & $\checkmark$ & Edge & Viable for $\alpha_2\lesssim2.2$ (Fig.~\ref{fig:bpl_family})\\
 \cline{2-11}
 & \multirow{3}{*}{Two populations, $\alpha=2.06$} & Poisson & $\checkmark$ & $\checkmark$ & $\checkmark$ & $\checkmark$ & $\times$ & $\times$ & -- & Timing fails\\
 & & $k\simeq0.42$ & $\checkmark$ & $\checkmark$ & $\checkmark$ & $\checkmark$ & $\checkmark$ & $\times$ & -- & Intervals too long\\
 & & On/off, $k\simeq0.5$ & $\checkmark$ & $\checkmark$ & $\checkmark$ & $\checkmark$ & $\checkmark$ & $\checkmark$ & $\checkmark$ & Viable, $f_{\rm nonrep}\lesssim0.95$ (Fig.~\ref{fig:two_population_family})\\
\enddata
\tablecomments{The first four diagnostic columns are the SCs, the next
two the burst timing, and the last the number of bursts per repeater
(Table~\ref{tab:observables}). The $\tau_{1/2}$ column is judged by the
KS probability of $C(\tau)$ at $\tau>1$~d. \checkmark = satisfied,
$\times$ = violated, Edge = satisfied only near the $2\sigma$ boundary,
-- = not evaluated because an earlier diagnostic already excludes the
model. All models use beam-weighted transits. Poisson bursts are Weibull
clustering with $k=1$. A shape $k$ alone denotes stationary Weibull
clustering, and on/off denotes Weibull clustering within active phases
of months to a year (\S~\ref{sec:onoff}). The checkmarks summarize
benchmark families at fixed $\alpha_1$ or $\alpha$, not a joint
likelihood fit.}
\end{deluxetable*}

\section{Discussion}
\label{sec:discussion}

Table~\ref{tab:model_summary} summarizes the constraints. Only
populations spanning $\mu\sim1$, with $\phi(r)\propto r^{-2}$ over the
probed rates, moderately clustered bursts, and active phases of months to a year, reproduce
all seven quantities.

\paragraph{Origin of $\alpha\simeq2$.} $\phi(r)$ is a distribution of
detected rates, $r=R\varepsilon$, where $R$ is the intrinsic rate and
$\varepsilon$ the fraction of bursts above threshold. For a burst energy
function $\propto E^{-\gamma}$ and sources uniform in Euclidean space,
the spread of distances alone gives a slope $\alpha_0=1+3/[2(\gamma-1)]$.
For the flat energy functions of the population, $\gamma\simeq1.3$--1.9
\citep{hashimoto22,shin23,gupta25}, $\alpha_0>2$, so the measured slope
must be intrinsic. One simple scenario gives it. Suppose each source
becomes active at some time and its rate then declines as
$r(t)\propto t^{-p}$, with sources activated at a rate
$\dot N(t)\propto t^q$. The active sources then follow
$\alpha=1+(1+q)/p$
\citep{lu20}. For a constant activation rate ($q=0$), $\alpha\simeq2$
gives $p\simeq1$, a rate that fades as the inverse of the source age.
The same framework gives the steeper tail of the BPL. A decay with an
early, fast phase ($p_1\simeq1$, matching $\alpha_1$) followed by a
slower one ($p_2\simeq0.5$--0.8 for $\alpha_2\simeq2.2$--3) gives a
break in $\phi(r)$ at a rate set by the break of the individual decay.
These are illustrations, not demonstrated mechanisms. For the
steep energy function of hyperactive repeaters, $\gamma\simeq2.5$
\citep{wu25}, $\alpha_0\simeq2$. If the distances of the sources then
spread their detected rates much more than their intrinsic rates differ,
$\phi(r)$ takes the slope $\alpha_0$ whatever the intrinsic rates, and
the measured slope reflects the distances rather than the sources.

\paragraph{Clustering.} The small same-transit fraction requires
moderate clustering, $k\simeq0.5$. Within single observations of a few
hours, the waiting times of active repeaters are close to Poisson,
$k\simeq1$ (\citealt{jahns23} for FRB~20121102A, and
\citealt{li26,wang25} for FRB~20240114A). Samples that span many days
give $k\simeq0.3$--0.9 \citep{opperman18,oostrum20,cruces21,zhang21,bhattacharyya26,konijn24},
and $k\simeq0.78$ for the full FAST sample of FRB~20240114A
\citep{li26}. Our $k$ describes the detected bursts. How it relates to
the shape $k'$ of the intrinsic bursts depends on the energy function,
through the detected fraction $\varepsilon$. Detection keeps a random
fraction $\varepsilon$ of the bursts, and Weibull clustering groups
bursts only within about one mean interval of the intrinsic rate. If
the flat energy function holds, $\varepsilon\gtrsim0.1$ for all sources
and much of the grouping survives. The detected $k\simeq0.5$ then
roughly equals $k'$, or $k'$ is somewhat smaller, and the comparison
with individual repeaters holds. If the steep energy function holds,
$\varepsilon\ll1$ for most sources. At most one burst of each group is
then detected, so the detected bursts look nearly Poisson for any $k'$.
The clustering must then come from storms of hours to a day that cover a
few to ten per cent of each active phase. A storm holds so many bursts
that several are detected even for $\varepsilon\ll1$. The detected $k$
then measures these storms rather than $k'$, and both readings
reproduce the timing (Appendix~\ref{app:monte_carlo}).

\paragraph{Long-term activity.} The first-repeat intervals of the Cat2 repeaters provide the first population-level evidence that repeaters are active only in phases of months to a year. These phases may recur with duty cycles $\lesssim0.2$, or a source may be active only once. The present time baseline cannot tell the two apart. Some individual repeaters show recurring phases. FRB~20121102A has shown long quiescent and highly active
stages \citep[e.g.,][]{li21}, and FRB~20201124A active stages of months
separated by quiescence of at least two years \citep{lanman22,xu22}. The
periodic windows of FRB~20180916B and FRB~20121102A, about 5~d every
16.3~d and 90~d every 157~d \citep{chime20_periodic,rajwade20,cruces21},
often attributed to precession or orbital motion
\citep[e.g.,][]{levin20,ioka20,wada21}, recur too often to produce this
quiescence and would be structure within the active phases. The
modulation may be intrinsic, as in the outbursts of Galactic magnetars
\citep{rea11,cotizelati18}, or due to propagation, such as plasma
lensing \citep[e.g.,][]{cordes17,chen24}. For the most active repeater,
FRB~20240114A, the energy budget of its persistent radio source limits
the duty cycle of its hyperactive phase, with up to $\sim700$ bursts per
hour, to $\lesssim10^{-2}$ \citep{matsumiya26}, which suggests that the
most intense activity is also the rarest. The population therefore
points to long-term activity changes as a common property of repeaters,
whatever their origin.

\paragraph{Telling the two resolutions apart.} The BPL and the
two-population model differ only through sources below the probed rates
(Figure~\ref{fig:bpl_family}b), and they predict different growth of
$Q\propto\Delta t^{\beta_r-\beta_s}$. In the BPL these sources are slow
repeaters. They begin to repeat once the sources near the break reach
$\mu\sim1$, at $\Delta t_b=\Delta t_{\max}/\epsilon$, about 50~yr of
CHIME for $\epsilon=0.1$. Before that the repeaters come from the
high-rate branch, and after it from the steep tail, which gives
$N_r\propto\Delta t^{\alpha_2-1}$ near $\mu\sim1$ and
$\propto\Delta t^{1+k}$ at $\mu\ll1$, whichever grows more slowly. The
singles stay $\propto\Delta t$ until even the slowest sources reach
$\mu\sim1$, at $\Delta t_c\equiv1/(\langle f\rangle_\delta r_{\min})$.
Hence, for the BPL,
\begin{equation}
Q_{\rm 1pop}\propto
\begin{cases}
\Delta t^{\alpha_1-2}, & \Delta t\ll\Delta t_b,\\[2pt]
\Delta t^{\min(\alpha_2-2,\,k)}, & \Delta t_b\ll\Delta t\ll\Delta t_c,
\end{cases}
\label{eq:q_slope_bpl}
\end{equation}
nearly constant now and rising later. In the two-population model the
one-off singles keep growing as $\Delta t$, and the repeaters grow as
$\Delta t^{\alpha-1}$ until the sources at the cutoff of the repeating
component reach $\mu\sim1$, at $\Delta t_c$ for that cutoff, and then stop
growing. Hence,
\begin{equation}
Q_{\rm 2pop}\propto
\begin{cases}
\Delta t^{\alpha-2}, & \Delta t\ll\Delta t_c,\\[2pt]
\Delta t^{-1}, & \Delta t\gg\Delta t_c.
\end{cases}
\label{eq:q_slope_two_population}
\end{equation}
A rising $Q$ would thus favor a single population with a steep tail, and
a constant or falling $Q$ with a linearly growing $N_s$ a one-off
component. Wide-field surveys with ten times the CHIME exposure per
source, such as BURSTT \citep{lin2022}, DSA/Chronoscope
\citep{connor2026}, and CHARTS \citep{charts1,charts2}, could reach this
regime within about 5~yr if the break lies near our $r_b$.

\paragraph{Comparison with previous studies.} \citet{james23} and \citet{hoffmann26}
explained Cat1 and Cat2 data with Poisson repeaters and a PL rate distribution,
with an index $-2.20^{+0.12}_{-0.22}$ consistent with ours.
\citet{beniamini25} matched the Cat1 repeater fraction with
$\alpha\simeq2.1$--2.3, and \citet{cook26} found with Cat2 that
50--100\% of the apparently one-off sources may be repeaters. These studies assume Poisson bursts, which our
burst timing excludes, and fit the counts at a single epoch. Clustering
adds repeaters at a given $\alpha$, so a Poisson fit maps the same data
onto a different rate distribution. 
\citet{lin24} found no repeats from ten FRBs detected in the
CHIME far sidelobes in Cat1 over 35,580~h, favoring either a
broad distribution of repetition rates or a genuinely one-off
population. In our framework, these sources could be slow
repeaters in the BPL or genuinely one-off sources. Using Cat2,
\citet{dong26} and \citet{sirota26} also suggested the existence of a genuinely
one-off population, based on the burst energy functions and
simulation-based inference, respectively. A genuinely one-off population remains allowed by our analysis.
A falling repeater fraction, as predicted by \citet{sirota26},
would favor such a population, whereas a rising fraction would
favor the BPL.
Our results also update Y24, which
fitted the SCs of a single Cat1 interval with a single PL of index 2.7
and $k\sim0.8$. Such a steep index across the probed rates would give
$\beta_r\simeq1.7$. It came mainly from a flattening of the single count
that lies within the scatter between intervals (Figure~\ref{fig:sc_measured}), and is better read as
the low-rate tail of our BPL. The conclusion of Y24 that most FRBs may
repeat remains allowed, but it is no longer required.

\section{Conclusions}
\label{sec:conclusions}

We measured how the numbers of single and repeating FRB sources in
CHIME/FRB Catalog 2 grow with observing time, and asked whether this
growth and the repeat bursts require a genuinely one-off population.
Our main conclusions are as follows.

\begin{itemize}

\item The exposure-corrected SCs, combined over all independent
intervals of each duration, grow linearly over $\sim300$--44,000~hr,
$N_s\propto\Delta t^{0.98\pm0.02}$ and $N_r\propto\Delta t^{1.06\pm0.14}$,
with no break. Their ratio stays at $Q=0.026^{+0.004}_{-0.006}$. A single
interval of the Cat1 length is not representative at
$\Delta t\lesssim10^3$~hr.

\item Of 77 repeaters, only 6 show their first two bursts within one
transit, the median first-repeat interval is 101~d, and 31 of the 52
repeaters with three or more bursts have five or more.

\item Populations whose sources all have $\mu\gg1$ are excluded because
their singles vanish. Populations whose sources all have $\mu\ll1$ are
excluded by the count--count slope for Poisson and weakly clustered bursts,
and by the first-repeat intervals and the number of bursts per repeater
for clustered ones. The sources must span $\mu\sim1$, with detected
rates distributed as $\phi(r)\propto r^{-\alpha}$, $\alpha\simeq2$, over
$\sim0.01$--1.5~hr$^{-1}$.

\item Below the probed rates the SCs fix only the number of sources. A
single PL extended to a finite cutoff, a tail that steepens below the
probed rates, and a genuinely one-off component with a fraction up to
0.95 all reproduce the data. The single PL needs a cutoff beyond the
active ages of magnetars unless $\alpha\gtrsim2.14$, whereas the other
two forms work for any $\alpha$ near 2.

\item The repeat bursts constrain the clustering. The few same-transit
first repeats require moderately clustered bursts, $k\simeq0.5$, and the long median interval requires active phases of months to a year. These may recur, with duty cycles $\lesssim0.2$ and cycles longer than about 5~yr, or occur only once in a finite lifetime.
The SCs are insensitive to this clustering, because $\phi(r)$ is close
to $r^{-2}$. The number of bursts per repeater mildly favors the flatter
members of the allowed models.

\item The BPL and the two-population model separate only when sources
near the break start to repeat. A rising repeater fraction would favor a
single population with a steep tail, and a linearly growing $N_s$ with a
constant or falling ratio a one-off component. Wide-field surveys with
ten times the CHIME exposure per source could reach this regime within
about 5~yr.

\end{itemize}

Together, our results link the SC evolution to the detected burst rates and temporal clustering of individual
sources, providing population-level constraints on FRB repeating
activity.

We note four possible caveats.
\begin{itemize}

\item The rates in $\phi(r)$ and the shape $k$ describe the detected
bursts. Whether they reflect the intrinsic rates and clustering depends
on the burst energy function of individual sources, which the present
data cannot determine (\S~\ref{sec:discussion}).

\item Our repeaters are those identified by CHIME. Counting the four
silver candidates of \citet{cook26} that keep two events after our cuts
as repeaters raises $Q$ by about 5\% and $\beta_r$ and $\Gamma_{\rm rs}$
by about 0.03, well within the errors.

\item CHIME may merge bursts close in time into one event. For FRB~20201124A,
sub-bursts 24--77~ms apart are catalogued as one burst, while bursts
0.03~s and more apart are distinct events \citep{lanman22}. Taking
$\sim0.1$~s as a rough limit, for $k=0.5$ this affects $\lesssim0.2$ of our six same-transit pairs, and a
source whose only repeat is merged changes $Q$ by $\lesssim0.3\%$.

\item Some apparently one-off sources may be episodic repeaters,
with short burst storms separated by long quiescent periods
\citep{mckinven26}. Their example, FRB~20210601A, is one of our six
same-transit repeaters. However, the median interval between
transits is about 100~d, requiring the active phase to last for
months (\S~\ref{sec:onoff}). Thus, such episodic activity does not
remove the need for long-lived source activity.

\end{itemize}

The apparent repeater fraction therefore does not directly measure the fraction of FRB sources that repeat. The present data allow both a single population
of repeaters, with detected rates distributed
approximately as $r^{-2}$ and moderately clustered bursts during
activity phases lasting months to years, and a mixed population that includes genuinely one-off sources. These scenarios make distinct
predictions for the future evolution of the SCs.
The growth of the counts over the coming years can tell the two apart.

\begin{acknowledgments}

We thank the CHIME/FRB team for making the second CHIME/FRB catalog publicly available, which is essential to this work. SY thanks Kunihito Ioka for useful discussions at the Yukawa Institute for Theoretical Physics (YITP) at Kyoto University, where early concepts of theoretical modeling in this work was initiated via YITP ``Atom-type Researcher" program in 2023. SY also thanks Ue-Li Pen for useful discussions on the distribution of repetition rates.
SY acknowledges the support from the National Science and Technology Council of Taiwan (NSTC) through grant numbers 113-2112-M-005-007-MY3 and 113-2811-M-005-006-. TH acknowledges the support from the NSTC through grants 113-2112-M-005-009-MY3, 113-2123-M-001-008-, 111-2112-M-005-018-MY3, and the Ministry of Education of Taiwan through a grant 113RD109.
TW acknowledges the support by JSPS KAKENHI Grant-in-Aid for Scientific Research
No. JP25KJ0024 and No. JP25K17378.
TG acknowledges the support of the NSTC through grants 108-2628-M007-004-MY3, 111-2112-M-007-021, and 112-2123-M-001-004-.
\end{acknowledgments}

\appendix
\renewcommand{\theHequation}{app.\theequation}

\section{Exposure}
\label{app:exposure}

\begin{figure}
\centering
\includegraphics[width=0.5\columnwidth]{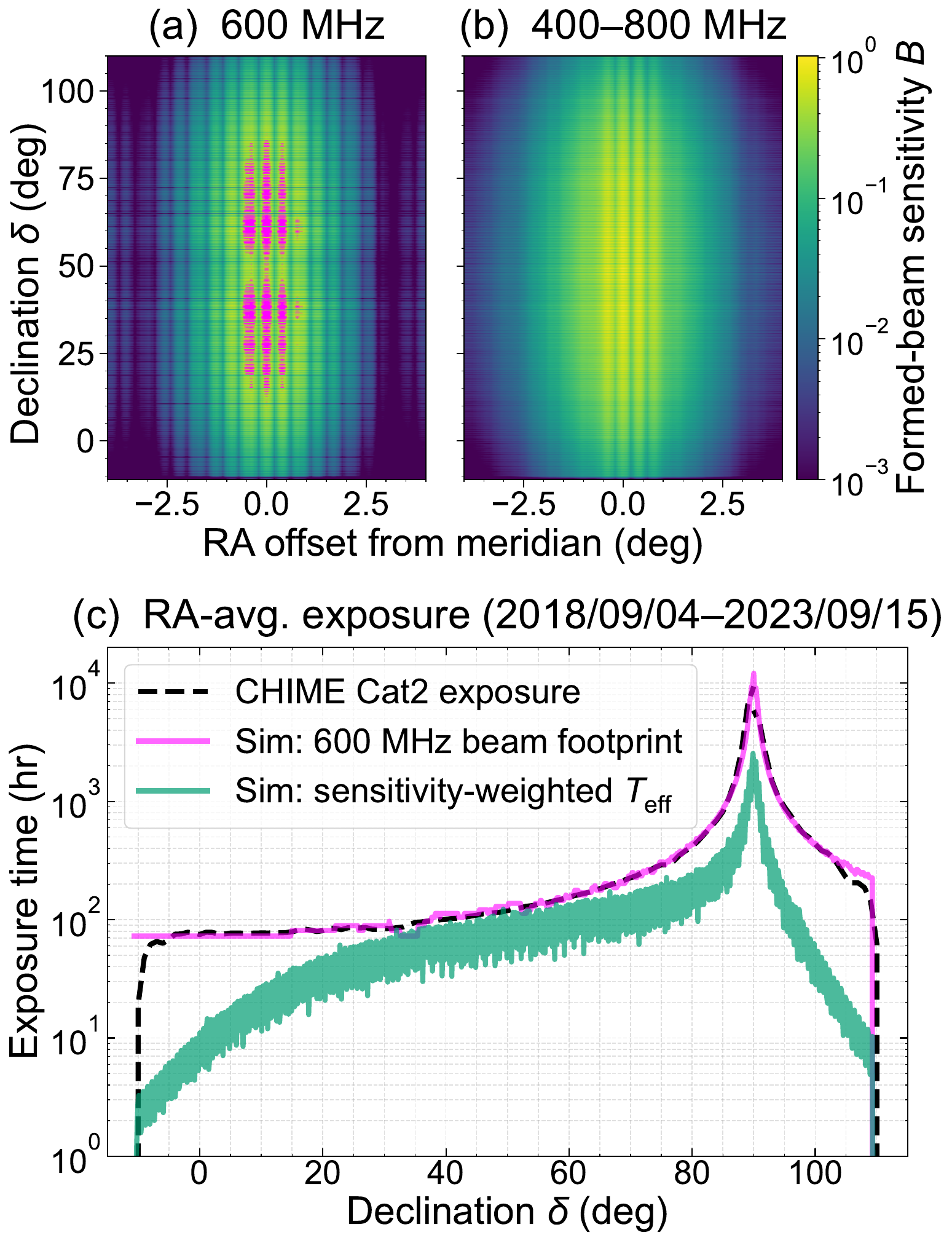}
\caption{(a) The 600~MHz formed-beam envelope. The magenta region is
above the FWHM sensitivity, which defines the Cat2 exposure. (b) The
formed-beam envelope averaged over 400--800~MHz, used in this study.
(c) Exposure as a function of declination. The dashed black curve is the
Cat2 exposure. Simulating the 600~MHz FWHM footprint reproduces it with
$T_{\rm survey}=1,837\times0.55$~d (magenta). The teal curve is the
sensitivity-weighted exposure $T_{\rm eff}(\delta)$ from the
frequency-averaged beam, with the same $T_{\rm survey}$.}
\label{fig:exposure}
\end{figure}

We use the public CHIME beam model, averaged over the XX and YY
polarizations and over 400--800~MHz \citep{james23}
(Figure~\ref{fig:exposure}b). For 1,000 declination bins we follow the
transit through the beam and compute $T_{\rm eff}(\delta)$. Upper and
lower transits are computed separately and combined for
$\delta\gtrsim70^\circ$. To fix the normalization, we repeat the
calculation with the 600~MHz FWHM footprint and adjust the effective
survey duration until it reproduces the Cat2 exposure. A duration of
$1,837\times0.55\simeq1,010$~d matches it (Figure~\ref{fig:exposure}c).
This only affects the normalization of the SCs. For the 3,157 sources,
$T_{\rm eff}$ has a mean of 185~hr. Its uncertainty follows from the
declination uncertainty of each source.

\section{Repeat probability of clustered bursts at low rate}
\label{app:low_rate_pairs}

\begin{figure}
\centering
\includegraphics[width=0.5\columnwidth]{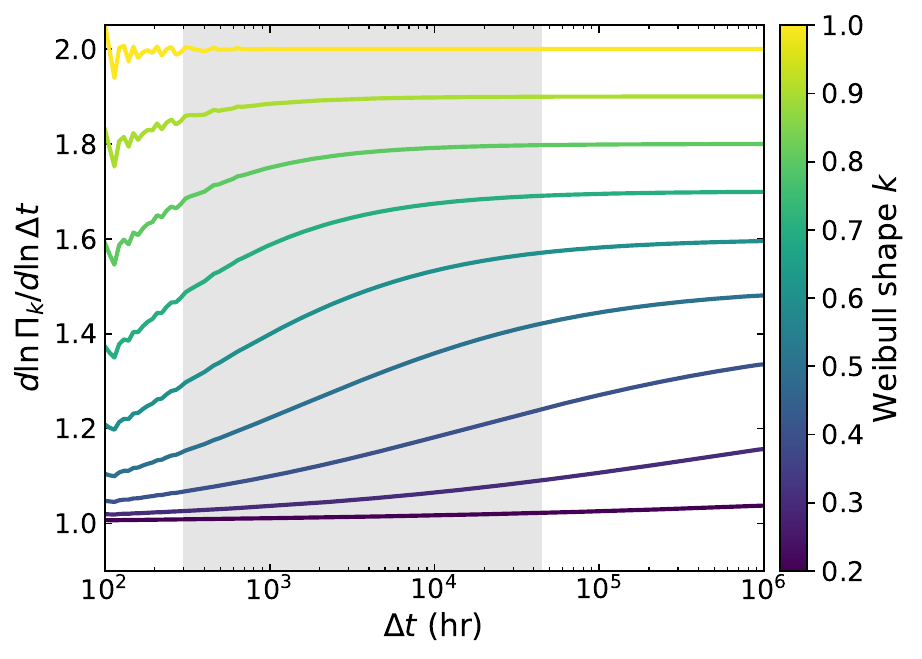}
\caption{Local slope of the pair factor, $d\ln\Pi_k/d\ln\Delta t$, for
Weibull clustering of shape $k$ (color) observed through the
beam-weighted daily transits. At low $\mu$ it equals the count--count slope.
The gray band marks the observed time baselines.}
\label{fig:pair_factor_slope}
\end{figure}

Let $g(\tau)$ be the pair correlation of the bursts. At low rate three
detections are much rarer than two, so the repeat probability equals the
expected number of pairs. For a set $W$ of observed instants,
\begin{equation}
P_{\geq2}^{\rm win}\simeq r^2\int_0^\infty A_W(\tau)\,g(\tau)\,d\tau,
\end{equation}
where $A_W(\tau)$ is the total length of pairs of observed instants
separated by $\tau$. For Weibull clustering of shape $k$,
$g(\tau)\simeq c_k(r\tau)^{k-1}$ for $r\tau\ll1$, with
$c_k=k\,\Gamma(1+1/k)^k$. In one window of length $T$ this gives
$P_{\geq2}\simeq\Gamma(1+1/k)^k\mu^{1+k}/(1+k)$, which is $\mu^2/2$ for
$k=1$. Through the daily transits, $N_r\propto\Pi_k\equiv\langle
f^{-1}\int A_W(\tau)\tau^{k-1}d\tau\rangle_\delta$. For transits of
duration $w=fP$ with period $P$,
\begin{equation}
f^{-1}\int A_W\tau^{k-1}d\tau\simeq\frac{\Delta t\,w^k+f\,\Delta t^{1+k}}{k(1+k)}.
\end{equation}
Pairs within one transit give the first term, which grows linearly, and
pairs in different transits the second. Figure~\ref{fig:pair_factor_slope}
shows the local slope of $\Pi_k$ from the beam map and the transit
sequence. At low $\mu$ it is the count--count slope. These expressions assume
$r\Delta t\ll1$. Higher rates make pairs in different transits more
Poisson-like and steepen $N_r$, so the limits on $k$ are conservative.

\section{The two-population family}
\label{app:two_population}

\begin{figure*}
\centering
\includegraphics[width=\textwidth]{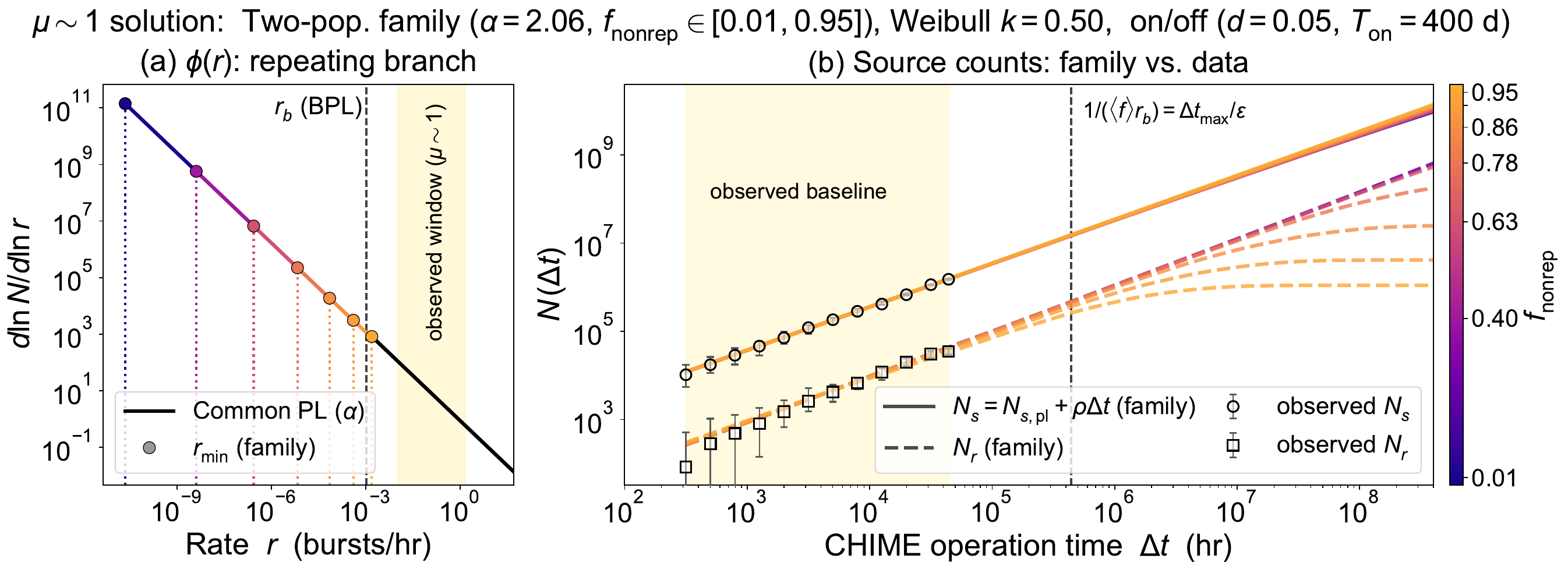}
\caption{Two-population family with a single PL ($\alpha=2.06$) for the
repeating component and a one-off component, for
$f_{\rm nonrep}=0.01$--0.95 (color), with the same on/off clustering as in
Figure~\ref{fig:bpl_family}. (a) Rate distribution of the repeating
component. Members share the PL and differ only in $r_{\min}$ (dots).
(b) $N_s$ (solid) and $N_r$ (dashed) compared with the data, and
extended to future time baselines.}
\label{fig:two_population_family}
\end{figure*}

The one-off component adds singles at a constant rate $\rho$, and
$f_{\rm nonrep}\equiv\rho\Delta t/N_s(\Delta t)$ at the reference
time baseline. The repeating component must then supply
$Q_{\rm obs}/(1-f_{\rm nonrep})$. Its ratio cannot fall below the
scale-free value $Q_{\rm floor}$, so solutions need
$f_{\rm nonrep}>1-Q_{\rm obs}/Q_{\rm floor}$. At $\alpha=2.06$ the floor
nearly vanishes, $Q_{\rm floor}\simeq2\times10^{-3}$, and there is no lower
bound. The larger $f_{\rm nonrep}$, the higher the cutoff of the
repeating component (Figure~\ref{fig:two_population_family}a). A
recurrence time within the active ages of magnetars needs
$f_{\rm nonrep}\gtrsim0.45$. For the largest fractions the cutoff
approaches the probed range and $N_r$ begins to level off, but even
$f_{\rm nonrep}=0.95$ gives $\beta_r=0.92$ and $\Gamma_{\rm rs}=0.94$,
within $1\sigma$. Unlike the BPL, $N_s$ grows linearly at long
time baselines.

\section{Monte Carlo evaluation of the windowed probabilities}
\label{app:monte_carlo}

Each declination node has an upper transit and, for
$\delta\geq70^\circ$, a lower transit half a day later, with detection
probability $p=B^{3/2}$ along the transit. For stationary clustering we
simulate the burst sequences of many sources at each rate and time baseline.
For each sequence we compute, exactly over the transit phase, the
probabilities of one detection, of at least two, and of at least two
with the first two in one transit.

For on/off activity, the active phases of a source are independent,
because the clustering restarts in each of them. For each rate we
tabulate the same probabilities for one active phase, as a function of
its length, from full burst sequences. At on-state rates
$r/d\geq2$~hr$^{-1}$ the bursts lose their correlation within a few mean
intervals, much less than the 12~hr between transits, and we tabulate
the probabilities for one transit instead. We then draw the active
phases of each source, with a random transit phase, and combine the
tables exactly. Every burst over the time baseline is thus included, even for
the fastest sources. The results agree with brute-force simulations
within the Monte Carlo errors, and with the exact probabilities for
Poisson bursts.

For the SCs we average over the nodes with weights $\cos\delta$ and
divide by $f$, as in Eq.~\eqref{eq:expected_counts}. At low $\mu$ we
use $P_1=\mu$. For the repeat bursts each source is weighted without
the factor $f^{-1}$, as in the catalog. We also tested two variants of
the representative model. A transit profile $p=B^{\gamma-1}$ with
$\gamma-1=0.3$--1.5 gives $F_{\rm same}=0.072$--0.082 and a median
interval shorter by at most 25\%, and leaves the SC slopes within 0.03.
Storms of 1--24~hr covering 3--10\% of each active phase, with Poisson
bursts inside, give $F_{\rm same}=0.05$--0.13 and $\tau_{1/2}=90$--110~d,
consistent with the catalog.

\bibliographystyle{aasjournal}

\end{document}